\documentclass[12pt]{article} 
\usepackage{epsf} 
\renewcommand{\thefootnote}{\fnsymbol{footnote}} 
 
\begin{document}

 
\begin{titlepage} 
\begin{flushright} 
\begin{tabular}{l} 
DESY 01--030\\ 
CERN--TH/2001--072\\ 
TTP01--07\\ 
hep--ph/0103121\\ 
March 2001 
\end{tabular} 
\end{flushright} 
 
\vspace*{1.8truecm} 
 
\begin{center} 
\boldmath 
{\Large \bf Exploring New Physics in the $B\to \phi K$ System} 
\unboldmath 
 
\vspace*{2.1cm} 
 
\smallskip 
\begin{center} 
{\sc {\large Robert Fleischer}}\footnote{E-mail: {\tt  
Robert.Fleischer@desy.de}} \\ 
\vspace*{2mm} 
{\sl Deutsches Elektronen-Synchrotron DESY, Notkestra\ss e 85,  
D--22607 Hamburg, Germany} 
\vspace*{1truecm}\\ 
{\sc{\large Thomas Mannel}}\footnote{E-mail: {\tt 
              Thomas.Mannel@cern.ch}}\\ 
\vspace*{2mm} 
{\sl CERN Theory Division, CH--1211 Geneva 23, Switzerland} 
\\ and \\ 
{\sl Institut f\"{u}r Theoretische Teilchenphysik, 
     Universit\"{a}t Karlsruhe, \\ D--76128 Karlsruhe, Germany} 
\end{center} 
 
\vspace{2.0truecm} 
 
{\large\bf Abstract\\[10pt]} \parbox[t]{\textwidth}{ 
Employing the $SU(2)$ isospin symmetry of strong interactions and  
estimates borrowed from effective field theory, we explore the  
impact of new physics on the decays $B^\pm\to \phi K^\pm$ and  
$B_d\to \phi K_{\rm S}$ in a model-independent manner. To this end,  
we introduce -- in addition to the usual mixing-induced CP asymmetry  
in $B_d\to \phi K_{\rm S}$ -- a set of three observables, which may  
not only provide ``smoking-gun'' signals for new-physics contributions  
to different isospin channels, but also valuable insights into hadron  
dynamics. Imposing dynamical hierarchies of amplitudes, we discuss  
various patterns of these observables, including also scenarios with  
small and large rescattering processes. Whereas the $B\to\phi K$  
system provides, in general, a powerful tool to search for indications  
of new physics, there is also an unfortunate case, where such effects  
cannot be distinguished from those of the Standard Model.  
} 
 
\vskip1.5cm 
 
\end{center} 
 
\end{titlepage}

\thispagestyle{empty} 
\vbox{} 
\newpage 
  
\setcounter{page}{1}

\setcounter{footnote}{0} 
\renewcommand{\thefootnote}{\arabic{footnote}}

\section{Introduction}\label{sec:intro} 
The experimental data collected at the $B$ factories will allow stringent 
tests of the Kobayashi--Maskawa picture of CP violation \cite{KM}.  
Among the various $B$ decays that can be used to achieve this goal  
\cite{revs}, $B\to \phi K$ decays play an important role. In these 
modes, which are governed by QCD penguin processes \cite{BphiK-old}, also  
electroweak (EW) penguins are sizeable \cite{BphiK-EW}, and physics beyond 
the Standard Model may have an important impact \cite{BphiK-NP}. In the  
summer of 2000, the observation of the $B^\pm\to \phi K^\pm$ channel was  
announced by the Belle and CLEO collaborations. The present results for 
the CP-averaged branching ratio are given as follows: 
\begin{equation}\label{EXP} 
\mbox{BR}(B^\pm\to \phi K^\pm)=\left\{\begin{array}{ll} 
\left(1.39^{+0.37+0.14}_{-0.33-0.24}\right) 
\times 10^{-5} & \mbox{(Belle \cite{belle})}\\ 
\left(5.5^{+2.1}_{-1.8}\pm0.6\right) 
\times 10^{-6} & \mbox{(CLEO \cite{cleo}).} 
\end{array} 
\right. 
\end{equation} 
The Belle and CLEO results are only marginally compatible with each other.  
Evidence for the neutral mode $B^0_d\to \phi K^0$ at the $2.9\sigma$ level, 
corresponding to a branching ratio of  
$\left(5.4^{+3.7}_{-2.7}\pm0.7\right)\times 10^{-6}$, was also reported by  
CLEO, whereas a significant signal for this decay has not yet been observed 
by the Belle collaboration. 
 
In our discussion of the $B\to \phi K$ system, we follow closely our  
recent $B\to J/\psi K$ analysis \cite{FM}, and make use of the $SU(2)$  
isospin symmetry of strong interactions to derive a model-independent  
parametrization of the $B^+\to \phi K^+$, $B^0_d\to \phi K^0$ decay  
amplitudes. After recapitulating the structure of the Standard-Model 
amplitudes, we include new-physics effects in a general manner, and  
estimate their generic size with the help of arguments borrowed from the 
picture of effective field theory. In order to deal with hadronic matrix  
elements, we impose certain dynamical hierarchies of decay amplitudes,  
where we distinguish between small and large rescattering effects. Although  
we do not consider the latter case, which is also not favoured by the  
``QCD factorization'' approach \cite{PQCD} and the present experimental  
upper bounds on $B\to K K$ branching ratios \cite{BF}, as a very  
likely scenario,\footnote{Arguments against this possibility, i.e.\ 
large rescattering effects, were also given in \cite{GIW}.} it deserves  
careful attention to separate possible new-physics effects from those of  
the Standard Model. Moreover, following the strategies proposed below,  
we may not only obtain insights into new physics, but also into hadron  
dynamics. To this end, we introduce -- in addition to the usual  
mixing-induced CP asymmetry in $B_d\to \phi K_{\rm S}$ -- a set of three  
observables, providing ``smoking-gun'' signals for new-physics contributions  
to different isospin channels. Two of these new-physics observables may be  
significantly enhanced by large rescattering processes. In general, the  
$B\to\phi K$ system offers powerful tools to search for new physics.  
However, there is also an unfortunate case, where such effects cannot  
be disentangled from those of the Standard Model. 
 
The outline of this paper is as follows: in Section~\ref{sec:pheno},  
we employ a low-energy effective Hamiltonian and the isospin 
symmetry of strong interactions to parametrize the $B^\pm\to \phi K^\pm$,  
$B_d\to \phi K_{\rm S}$ decay amplitudes arising within the Standard  
Model. The impact of new physics on these amplitudes is explored in a  
model-independent way in Section~\ref{sec:NP}, where we make use of  
dimensional estimates following from effective field theory, and  
introduce plausible dynamical hierarchies of amplitudes. The set of  
observables to search for ``smoking-gun'' signals of new-physics  
contributions to different isospin channels of the $B\to \phi K$  
decay amplitudes is introduced in Section~\ref{sec:obs}, and is discussed  
in further detail in Section~\ref{sec:disc}. In Section~\ref{sec:concl},  
our conclusions are summarized.

\boldmath 
\section{Phenomenology of $B\to \phi K$ Decays}\label{sec:pheno} 
\unboldmath 
The $B\to \phi K$ system is described by the following low-energy effective  
Hamiltonian: 
\begin{equation}\label{Heff} 
{\cal H}_{\rm eff}=\frac{G_{\rm F}}{\sqrt{2}}\left[V_{cs}V_{cb}^\ast 
\left({\cal Q}_{\rm CC}^c-{\cal Q}_{\rm QCD}^{\rm pen}- 
{\cal Q}_{\rm EW}^{\rm pen}\right)+V_{us}V_{ub}^\ast 
\left({\cal Q}_{\rm CC}^u-{\cal Q}_{\rm QCD}^{\rm pen}- 
{\cal Q}_{\rm EW}^{\rm pen}\right)\right], 
\end{equation} 
where the ${\cal Q}$ are linear combinations of perturbative Wilson  
coefficient functions and four-quark operators, 
consisting of current--current  
(CC), QCD penguin and EW penguin operators. As discussed in \cite{FM},  
this Hamiltonian is a combination of isospin $I=0$ and $I=1$  
pieces: 
\begin{equation}\label{ham-decom} 
{\cal H}_{\rm eff} = {\cal H}_{\rm eff}^{I=0} + {\cal H}_{\rm eff}^{I=1}, 
\end{equation} 
where ${\cal H}_{\rm eff}^{I=0}$ receives contributions from all of 
the operators appearing in (\ref{Heff}), whereas ${\cal H}_{\rm eff}^{I=1}$  
is due to only ${\cal Q}_{\rm CC}^u$ and ${\cal Q}_{\rm EW}^{\rm pen}$.  
If we employ the $SU(2)$ isospin flavour symmetry of strong interactions,  
we obtain 
\begin{eqnarray} 
\langle \phi K^+|{\cal H}_{\rm eff}^{I=0}|B^+\rangle&=& 
+\langle \phi K^0|{\cal H}_{\rm eff}^{I=0}|B^0_d\rangle\label{iso1}\\ 
\langle \phi K^+|{\cal H}_{\rm eff}^{I=1}|B^+\rangle&=& 
-\langle \phi K^0|{\cal H}_{\rm eff}^{I=1}|B^0_d\rangle,\label{iso2} 
\end{eqnarray} 
yielding 
\begin{equation}\label{AMPLp1} 
A(B^+\to \phi K^+)=\frac{G_{\rm   
F}}{\sqrt{2}}\left[V_{cs}V_{cb}^\ast\left\{ 
{\cal A}_c^{(0)}-{\cal A}_c^{(1)}\right\}+V_{us}V_{ub}^\ast\left\{{\cal  
A}_u^{(0)}-{\cal A}_u^{(1)}\right\}\right] 
\end{equation} 
\begin{equation}\label{AMPLd1} 
A(B^0_d\to \phi K^0)=\frac{G_{\rm F}}{\sqrt{2}}\left[V_{cs}V_{cb}^\ast 
\left\{{\cal A}_c^{(0)}+{\cal A}_c^{(1)}\right\}+V_{us}V_{ub}^\ast 
\left\{{\cal A}_u^{(0)}+{\cal A}_u^{(1)}\right\}\right], 
\end{equation} 
where the CP-conserving strong amplitudes\footnote{The labels of 
${\cal A}^{(0)}$ and ${\cal A}^{(1)}$ refer to the isospin channels 
$I=0$ and $I=1$, respectively.} 
\begin{equation}\label{ampl-c} 
{\cal A}_c^{(0)}={\cal A}_{\rm CC}^c-{\cal A}_{\rm QCD}^{\rm pen}- 
{\cal A}_{\rm EW}^{(0)},\quad 
{\cal A}_c^{(1)}=-{\cal A}_{\rm EW}^{(1)} 
\end{equation} 
\begin{equation}\label{ampl-u} 
{\cal A}_u^{(0)}={\cal A}_{\rm CC}^{u (0)}-{\cal A}_{\rm QCD}^{\rm pen}- 
{\cal A}_{\rm EW}^{(0)},\quad 
{\cal A}_u^{(1)}={\cal A}_{\rm CC}^{u (1)}-{\cal A}_{\rm EW}^{(1)} 
\end{equation} 
can be expressed in terms of hadronic matrix elements  
$\langle \phi K|{\cal Q}|B\rangle$. Taking into account that 
\begin{equation} 
V_{cs}V_{cb}^\ast=\left(1-\frac{\lambda^2}{2}\right)\lambda^2A,\quad 
V_{us}V_{ub}^\ast=\lambda^4 A\, R_b\, e^{i\gamma}, 
\end{equation} 
where $\gamma$ is the usual angle of the unitarity triangle of the CKM  
matrix \cite{revs}, and  
\begin{equation} 
\lambda\equiv|V_{us}|=0.22, \quad A\equiv|V_{cb}|/\lambda^2=0.81\pm0.06,  
\quad R_b\equiv|V_{ub}/(\lambda V_{cb})|=0.41\pm0.07, 
\end{equation} 
we finally arrive at 
\begin{equation}\label{AMPLp2} 
A(B^+\to \phi K^+)=\frac{G_{\rm F}}{\sqrt{2}} 
\left(1-\frac{\lambda^2}{2}\right)\lambda^2A\left\{{\cal A}_c^{(0)}- 
{\cal A}_c^{(1)}\right\}\left[1+\frac{\lambda^2 R_b}{1-\lambda^2/2} 
\left\{\frac{{\cal A}_u^{(0)}-{\cal A}_u^{(1)}}{{\cal A}_c^{(0)}- 
{\cal A}_c^{(1)}}\right\}e^{i\gamma}\right] 
\end{equation} 
\begin{equation}\label{AMPLd2} 
A(B^0_d\to \phi K^0)=\frac{G_{\rm F}}{\sqrt{2}} 
\left(1-\frac{\lambda^2}{2}\right)\lambda^2A\left\{{\cal A}_c^{(0)}+ 
{\cal A}_c^{(1)}\right\}\left[1+\frac{\lambda^2 R_b}{1-\lambda^2/2} 
\left\{\frac{{\cal A}_u^{(0)}+{\cal A}_u^{(1)}}{{\cal A}_c^{(0)}+ 
{\cal A}_c^{(1)}}\right\}e^{i\gamma}\right]. 
\end{equation} 
 
\begin{figure} 
\begin{center} 
\leavevmode 
\epsfysize=4.8truecm  
\epsffile{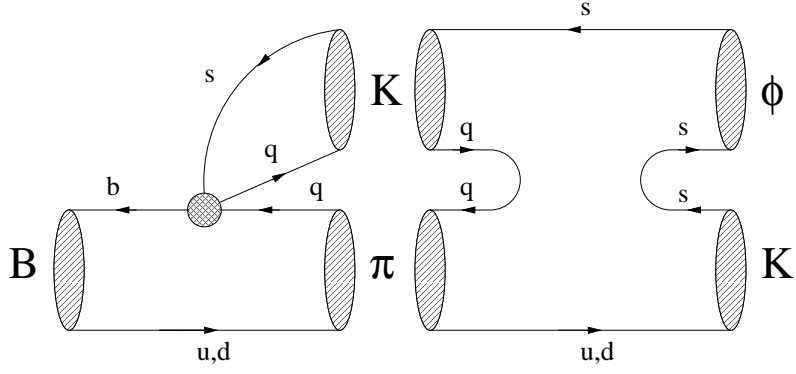} 
\end{center} 
\vspace*{-0.4truecm} 
\caption{Rescattering processes contributing to $B\to\phi K$ through  
penguin-like topologies with internal $q$-quark exchanges ($q\in\{u,d\}$).  
The shaded circle represents insertions of the corresponding current--current  
operators.}\label{fig:rescatter1} 
\end{figure}

\begin{figure} 
\vspace*{0.6truecm} 
\begin{center} 
\leavevmode 
\epsfysize=4.4truecm  
\epsffile{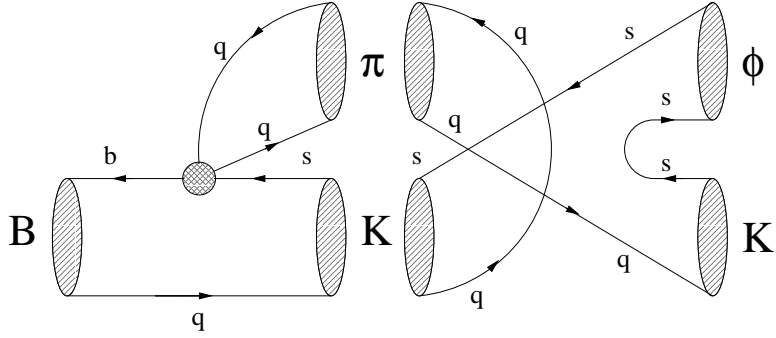} 
\end{center} 
\vspace*{-0.4truecm} 
\caption{Rescattering processes contributing to $B\to\phi K$  
through annihilation topologies. The shaded circle represents  
insertions of current--current operators  
($q\in\{u,d\}$).}\label{fig:rescatter2} 
\end{figure}

At first sight, expressions (\ref{AMPLp2}) and (\ref{AMPLd2}) are 
completely analogous to the ones for the $B^+\to J/\psi K^+$ and  
$B^0_d\to J/\psi K^0$ amplitudes given in \cite{FM}. However, the  
dynamics, which is encoded in the strong amplitudes ${\cal A}$, is  
very different. In particular, the current--current operators  
${\cal Q}_{\rm CC}^c$ cannot contribute to $B\to \phi K$ decays, i.e.\ 
to ${\cal A}_{\rm CC}^c$, through tree-diagram-like topologies; they may  
only do so through penguin topologies with internal charm-quark  
exchanges, which include also  
\begin{equation}\label{rescatter-c} 
B^+\to\{D_s^+\overline{D^0},...\}\to\phi K^+,\quad  
B^0_d\to\{D_s^+D^-,...\}\to\phi K^0 
\end{equation} 
rescattering processes \cite{BFM}, and may actually play an important  
role \cite{pens}. On the other hand, the ${\cal A}_{\rm CC}^{u (0,1)}$  
amplitudes receive contributions from penguin processes with internal  
up- and down-quark exchanges, as well as from annihilation  
topologies.\footnote{Note that the isospin projection operators  
${\cal Q}\sim(\overline{u}u\pm\overline{d}d)(\overline{b}s)$ involve  
also $\overline{d}d$ quark currents.} Such penguins may also be  
important, in particular in the presence of large rescattering processes  
\cite{BFM,FSI}; a similar comment applies to annihilation topologies. In  
the $B\to\phi K$ system, the relevant rescattering processes are  
\begin{equation}\label{rescatter-u} 
B^+\to\{K^+\pi^0,...\}\to\phi K^+,\quad  
B^0_d\to\{K^+\pi^-,...\}\to\phi K^0, 
\end{equation} 
containing -- in addition to long-distance penguins -- also annihilation 
processes (see Figs.~\ref{fig:rescatter1} and \ref{fig:rescatter2}). In 
contrast to (\ref{rescatter-c}), large rescattering effects of the kind 
described by (\ref{rescatter-u}) may affect the search for new physics 
with $B\to \phi K$ decays, since these processes are associated with the 
weak phase factor $e^{i\gamma}$. Moreover, they involve ``light'' 
intermediate states, and are hence expected to be enhanced more easily, 
dynamically, through long-distance effects than (\ref{rescatter-c}), 
which involve ``heavy'' intermediate states.  
  
As is well known, the $\phi$-meson is an almost pure $\overline{s}s$ state;  
the mixing angle with its isoscalar partner $\omega\sim (\overline{u}u+ 
\overline{d}d)/\sqrt{2}$ is small, i.e.\ at the few per cent level. Whereas  
$\omega$--$\phi$ mixing does not at all affect the isospin relations  
(\ref{iso1}) and (\ref{iso2}), which rely on the fact that the $\phi$ is  
an isospin singlet, it has an impact on the size of the amplitudes  
${\cal A}_{\rm CC}^{u (0,1)}$, since an $\omega$ component of the $\phi$  
state permits current--current operator contributions through  
tree-diagram-like topologies. However, the arguments given below are not  
modified by the small $\omega$--$\phi$ mixing. 
 
Let us now have a closer look at the structure of the $B\to \phi K$ 
decay amplitudes, focusing first on the case corresponding to small  
rescattering effects. Looking at (\ref{ampl-c}) and (\ref{ampl-u}),  
we expect 
\begin{equation}\label{hier1} 
\left|{\cal A}_u^{(0,1)}/{\cal A}_c^{(0)}\right|={\cal O}(1). 
\end{equation} 
In the case of the amplitude ${\cal A}_c^{(1)}$, the situation is different. 
Here we have to deal with an amplitude that is essentially due to  
EW penguins. Moreover, the $B\to \phi K$ matrix elements of $I=1$ operators,  
having the general flavour structure 
\begin{equation}\label{I1-ops} 
{\cal Q}_{I=1}\sim (\overline{u}u-\overline{d}d)(\overline{b}s),  
\end{equation} 
are expected to suffer from a dynamical suppression. In order to keep track 
of these features, we introduce, as in \cite{FM}, a ``generic'' expansion  
parameter $\overline{\lambda}={\cal O}(0.2)$ \cite{hierarchy}, which is of  
the same order as the Wolfenstein parameter $\lambda=0.22$, and suggests 
\begin{equation}\label{hier2} 
\left|{\cal A}_c^{(1)}/{\cal A}_c^{(0)}\right|= 
\underbrace{{\cal O}(\overline{\lambda})}_{{\rm EW\, penguins}} 
\times\underbrace{{\cal O}(\overline{\lambda})}_{{\rm Dynamics}} 
={\cal O}(\overline{\lambda}^2). 
\end{equation} 
Consequently, we obtain 
\begin{equation}\label{SM-ampl} 
A(B^+\to \phi K^+)={\cal A}_{\rm SM}^{(0)} 
\left[1+{\cal O}(\overline{\lambda}^2)\right]=A(B^0_d\to \phi K^0), 
\end{equation} 
with  
\begin{equation}\label{ASM0} 
{\cal A}_{\rm SM}^{(0)}\equiv  
\frac{G_{\rm F}}{\sqrt{2}}\lambda^2A\,{\cal A}_c^{(0)}. 
\end{equation} 
The terms entering (\ref{SM-ampl}) at the $\overline{\lambda}^2$ level 
contain also pieces that are proportional to the weak-phase factor 
$e^{i\gamma}$,  thereby leading to direct CP violation in the 
$B\to \phi K$ system.

Let us now consider large rescattering effects of the kind given in 
(\ref{rescatter-u}). In the worst case, (\ref{hier1}) would be dynamically  
enhanced as  
\begin{equation}\label{hier1-res} 
\left|{\cal A}_u^{(0,1)}/{\cal A}_c^{(0)}\right|= 
{\cal O}(1/\overline{\lambda}), 
\end{equation} 
and the dynamical suppression in (\ref{hier2}) would no longer be  
effective, i.e. 
\begin{equation}\label{hier2-res} 
\left|{\cal A}_c^{(1)}/{\cal A}_c^{(0)}\right|={\cal O}(\overline{\lambda}). 
\end{equation} 
In such a scenario, (\ref{SM-ampl}) would receive corrections of  
${\cal O}(\overline{\lambda})$, involving also $e^{i\gamma}$. This  
feature may complicate the search for new physics with the help of  
CP-violating effects in $B\to\phi K$ decays. On the other hand, the  
rescattering processes described by (\ref{rescatter-c}) may only affect  
the amplitude ${\cal A}_{\rm SM}^{(0)}$ sizeably through its  
${\cal A}_{\rm CC}^c$ piece, and are not related to a CP-violating 
weak-phase factor within the Standard Model.  
 
Before turning to new physics, we would like to point out an interesting  
difference between the $B\to \phi K$ system and the $B\to J/\psi K$ decays  
discussed in \cite{FM}. In the $B\to J/\psi K$ case, the Standard-Model  
amplitudes corresponding to (\ref{SM-ampl}) receive corrections at the  
$\overline{\lambda}^3$ level, which may be enhanced to  
${\cal O}(\overline{\lambda}^2)$ in the presence of large 
rescattering effects. Consequently, within the Standard Model, there may  
be direct CP-violating effects in $B\to J/\psi K$ transitions of at most  
${\cal O}(\overline{\lambda}^2)$, whereas such asymmetries may already 
arise at the $\overline{\lambda}$ level in the $B\to \phi K$ system.  
On the other hand, as $B\to \phi K$ modes are governed by penguin processes, 
their decay amplitudes are more sensitive to new physics. In the next  
section, we have a closer look at the generic size of such effects, using  
dimensional arguments inspired by effective field theory.

\section{Effects of Physics Beyond the Standard Model}\label{sec:NP} 
The most general way of introducing new physics is to employ the picture  
of effective field theory, where we write down all possible dim-6  
operators and construct the generalization of the Standard-Model effective  
Hamiltonian (\ref{Heff}) at the scale of the $b$ quark. The relevant  
operators are again dim-6 operators, where the Wilson coefficients of 
those already present in the Standard Model now contain also a possible  
piece of new physics. The problem with this generic point of view is that  
the list of possible dim-6 operators contains close to one hundred entries  
\cite{dim-6}, not yet taking into account the flavour structure, making  
this general approach almost useless. However, as we are dealing with  
non-leptonic decays in which, because of our ignorance of the hadronic 
matrix elements, there is no senistivity neither to the helicity structure 
of the operators nor to their colour structure, only the flavour structure  
of the operators is relevant for us.  
 
A discussion of the $\Delta B = \pm 2$ operators mediating  
$B^0_d$--$\overline{B^0_d}$ mixing within this framework was given  
in \cite{FM}. For a characteristic new-physics scale $\Lambda$ in the  
TeV regime, new-physics contributions could in principle be as large as  
those of the Standard Model. This well-known feature was also found in  
several model-dependent studies of physics beyond the Standard Model  
\cite{NP-revs}. As far as CP violation is concerned, new-physics  
contributions involving also new CP-violating phases are of particular  
interest. In this case, not only the ``strength'' of  
$B^0_d$--$\overline{B^0_d}$ mixing is affected, which would manifest  
itself as an inconsistency in the usual ``standard analysis'' of the  
unitarity triangle \cite{AL}, but also ``mixing-induced'' CP-violating  
asymmetries \cite{revs}, arising, for instance, in $B_d\to J/\psi K_{\rm S}$  
or $B_d\to \phi K_{\rm S}$ decays.  
 
Let us now turn to the $B\to \phi K$ amplitudes. As in (\ref{ham-decom}), 
also in the presence of new physics, the corresponding low-energy effective  
Hamiltonian can be decomposed into $I=0$ and $I=1$ pieces, where the new  
physics may affect the Wilson coefficients and may introduce new  
dim-6 operators, thereby modifying (\ref{SM-ampl}) as follows: 
\begin{eqnarray} 
A(B^+\to \phi K^+)&=&{\cal A}_{\rm SM}^{(0)} 
\left[1+\sum_k v_0^{(k)}e^{i\Delta_0^{(k)}}e^{i\Phi_0^{(k)}}- 
\sum_j v_1^{(j)}e^{i\Delta_1^{(j)}}e^{i\Phi_1^{(j)}} 
\right]\label{ampl-NPp}\\ 
A(B^0_d\to \phi K^0)&=&{\cal A}_{\rm SM}^{(0)} 
\left[1+\sum_k v_0^{(k)}e^{i\Delta_0^{(k)}}e^{i\Phi_0^{(k)}}+ 
\sum_j v_1^{(j)}e^{i\Delta_1^{(j)}}e^{i\Phi_1^{(j)}}\right].\label{ampl-NPd}  
\end{eqnarray} 
Here $v_0^{(k)}$ and $v_1^{(j)}$ correspond to the $I=0$ and $I=1$ pieces,  
respectively, $\Delta_{0}^{(k)}$ and $\Delta_{1}^{(j)}$ are CP-conserving  
strong phases, and $\Phi_{0}^{(k)}$ and $\Phi_{1}^{(j)}$ the corresponding  
CP-violating weak phases. The amplitudes for the CP-conjugate processes  
can be obtained straightforwardly from (\ref{ampl-NPp}) and  
(\ref{ampl-NPd}) by reversing the signs of the weak phases. The labels  
$k$ and $j$ distinguish between different new-physics contributions to the   
$I=0$ and $I=1$ sectors. 
 
As we have already noted, within the Standard Model, $B\to \phi K$  
decays are governed by QCD penguins. Neglecting, for simplicity,  
EW penguins and the proper renormalization-group evolution, we may  
write 
\begin{equation}\label{SM-pen-expr} 
{\cal A}^{(0)}_{\rm SM}\sim\frac{G_{\rm F}}{\sqrt{2}}\lambda^2 A  
\left[\frac{\alpha_s}{4\pi}\, {\cal C}\right] \langle P_{\rm QCD}\rangle, 
\end{equation} 
where ${\cal C}={\cal O}(1)$ is a perturbative short-distance coefficient, 
which is multiplied by the characteristic loop factor $\alpha_s/(4\pi)$,  
and $P_{\rm QCD}$ denotes an appropriate linear combination of QCD penguin  
operators \cite{BF-rev}. Since (\ref{SM-pen-expr}) is a doubly  
Cabibbo-suppressed loop amplitude, new physics could well be of the same  
order of magnitude. If we assume that the physics beyond the Standard Model  
is associated with a scale $\Lambda$ and impose that it yields contributions  
to the $B\to \phi K$ amplitudes of the same size as the Standard Model, we  
obtain 
\begin{equation}\label{generic} 
\frac{G_{\rm F}}{\sqrt{2}}\frac{M_{W}^2}{\Lambda^2}\sim  
\frac{G_{\rm F}}{\sqrt{2}}\lambda^2 A \left[\frac{\alpha_s}{4\pi}\, {\cal C} 
\right], 
\end{equation} 
corresponding to $\Lambda\sim 3$\,TeV.\footnote{In our numerical  
estimate, we have assumed $A \times {\cal C}\sim1$ and  
$\alpha_s=\alpha_s(m_b)\sim 0.2$.}\,  
Consequently, for a generic new-physics scale in the TeV regime, which  
was also considered in our $B\to J/\psi K$ analysis \cite{FM}, we may have 
\begin{equation} 
v_0^{(k)}={\cal O}(1).  
\end{equation} 
In the case where the new-physics effects are less pronounced, it may 
be difficult to disentangle them from the Standard-Model contributions. 
We shall come back to this issue in Section~\ref{sec:disc}, where we 
shall discuss various scenarios.  
 
Concerning possible new-physics contributions to the $I=1$ sector, we  
assume a ``generic strength'' of the corresponding operators similar 
to (\ref{generic}). However, since the $I=1$ operators have the general  
flavour structure given in (\ref{I1-ops}), they are expected to suffer from  
a dynamical suppression in $B\to\phi K$ decays. As in (\ref{hier2}), we  
assume that this brings a factor of $\overline{\lambda}$ into the game,  
yielding 
\begin{equation}\label{gen-strength1} 
v_1^{(j)}={\cal O}(\overline{\lambda}). 
\end{equation} 
If we impose such a hierarchy of amplitudes, the new-physics contributions  
to the $I=1$ sector would be enhanced by a factor of  
${\cal O}(\overline{\lambda})$ with respect to the $I=1$ Standard-Model  
pieces. This may actually be the case if new physics shows up, for example,  
in EW penguin processes.  
 
Consequently, we finally arrive at  
\begin{equation}\label{ampl-hier1} 
A(B\to \phi K)={\cal A}_{\rm SM}^{(0)}\biggl[1+ 
\underbrace{{\cal O}(1)}_{{\rm NP}_{I=0}}+ 
\underbrace{{\cal O}(\overline{\lambda})}_{{\rm NP}_{I=1}}+ 
\underbrace{{\cal O}(\overline{\lambda}^2)}_{{\rm SM}}\biggr]. 
\end{equation} 
In deriving this expression, we have assumed that the $B\to \phi K$ decays  
are not affected by rescattering effects. On the other hand, in the presence  
of large rescattering processes of the kind described by (\ref{rescatter-u}),  
the dynamical suppression assumed in (\ref{gen-strength1}) would no longer  
be effective, thereby yielding $v_1^{(j)}={\cal O}(1)$. Analogously, the 
$B\to \phi K$ matrix elements of $I=0$ operators with flavour structure 
\begin{equation}\label{I0-ops1} 
{\cal Q}_{I=0}^{\overline{u}u,\overline{d}d} 
\sim (\overline{u}u+\overline{d}d)(\overline{b}s) 
\end{equation} 
would no longer be suppressed with respect to those of the dynamically  
favoured $I=0$ operators  
\begin{equation}\label{I0-ops2} 
{\cal Q}_{I=0}^{\overline{s}s}\sim (\overline{s}s)(\overline{b}s), 
\end{equation} 
and would also contribute to $v_0^{(k)}$ at ${\cal O}(1)$. A similar  
comment applies to the matrix elements of the $I=0$ operators with 
the following flavour content: 
\begin{equation}\label{I0-ops3} 
{\cal Q}_{I=0}^{\overline{c}c}\sim (\overline{c}c)(\overline{b}s), 
\end{equation} 
whose dynamical suppression in $B\to\phi K$ decays may be reduced through 
rescattering effects of the kind given in (\ref{rescatter-c}), which may 
also affect the ${\cal A}_{\rm SM}^{(0)}$ amplitude, as we have noted 
above. Consequently, in the presence of large rescattering effects,  
the decay amplitude (\ref{ampl-hier1}) is modified as follows: 
\begin{equation}\label{ampl-hier2} 
\left.A(B\to \phi K)\right|_{\rm res.}= 
\left.{\cal A}_{\rm SM}^{(0)}\right|_{\rm res.}\times 
\biggl[1+\underbrace{{\cal O}(1)}_{{\rm NP}_{I=0}}+ 
\underbrace{{\cal O}(1)}_{{\rm NP}_{I=1}}+ 
\underbrace{{\cal O}(\overline{\lambda})}_{{\rm SM}}\biggr]. 
\end{equation} 
Let us emphasize once again that the rescattering contributions to the  
prefactor on the right-hand side of this equation are due to  
(\ref{rescatter-c}), whereas the hierarchy in square brackets is  
governed by large rescattering processes of the type described  
by (\ref{rescatter-u}). 
 
In the following section, we introduce a set of observables, allowing 
us to separate the $I=0$ contributions from the $I=1$ sector. These 
observables play a key role to search for new physics and to obtain, 
moreover, valuable insights into the $B\to \phi K$ hadron dynamics.

\boldmath 
\section{Observables of $B\to \phi K$ Decays}\label{sec:obs} 
\unboldmath 
The decays $B^+\to \phi K^+$, $B^0_d\to \phi K^0$ and their 
charge conjugates are described by four decay amplitudes $A_i$. 
If the corresponding rates are measured, the $|A_i|^2$ can be 
extracted. Since we are not interested in the overall normalization  
${\cal A}_{\rm SM}^{(0)}$ of these amplitudes, we may construct the  
following three independent observables with the help of the $|A_i|^2$: 
\begin{equation}\label{def-AP} 
{\cal A}_{\rm CP}^{(+)}\equiv 
\frac{|A(B^+\to \phi K^+)|^2-|A(B^-\to \phi K^-)|^2}{|A(B^+\to  
\phi K^+)|^2+|A(B^-\to \phi K^-)|^2} 
\end{equation} 
\begin{equation}\label{def-A0} 
{\cal A}_{\rm CP}^{{\rm dir}}\equiv 
\frac{|A(B^0_d\to \phi K^0)|^2-|A(\overline{B^0_d}\to \phi 
\overline{K^0})|^2}{|A(B^0_d\to \phi K^0)|^2+ 
|A(\overline{B^0_d}\to \phi\overline{K^0})|^2} 
\end{equation} 
\begin{equation}\label{def-A} 
{\cal B} \equiv\frac{\langle|A(B_d\to \phi K)|^2\rangle- 
\langle|A(B^\pm\to \phi K^\pm)|^2\rangle}{\langle|A(B_d\to  
\phi K)|^2\rangle+\langle|A(B^\pm\to \phi K^\pm)|^2\rangle}, 
\end{equation} 
where the ``CP-averaged'' amplitudes are defined in the usual way: 
\begin{eqnarray} 
\langle|A(B_d\to \phi K)|^2\rangle&\equiv& 
\frac{1}{2}\left[|A(B^0_d\to \phi K^0)|^2 
+|A(\overline{B^0_d}\to \phi \overline{K^0})|^2\right]\\ 
\langle|A(B^\pm\to \phi K^\pm)|^2\rangle&\equiv& 
\frac{1}{2}\left[|A(B^+\to \phi K^+)|^2+|A(B^-\to \phi K^-)|^2\right]. 
\end{eqnarray} 
It should be noted that ${\cal B}$ is a CP-conserving quantity. In the  
case of the neutral decay $B_d\to \phi K_{\rm S}$, interference between  
$B^0_d$--$\overline{B^0_d}$ mixing and decay processes yields an additional  
observable \cite{revs}: 
\begin{equation}\label{time-dep} 
\frac{\Gamma(B^0_d(t)\to \phi K_{\rm S})- 
\Gamma(\overline{B^0_d}(t)\to \phi K_{\rm S})}{\Gamma(B^0_d(t)\to  
\phi K_{\rm S})+\Gamma(\overline{B^0_d}(t)\to \phi K_{\rm S})} 
={\cal A}_{\rm CP}^{{\rm dir}}\,\cos(\Delta M_d t)+ 
{\cal A}_{\rm CP}^{\rm mix}\,\sin(\Delta M_d t), 
\end{equation} 
where the ``direct'' CP-violating contribution  
${\cal A}_{\rm CP}^{{\rm dir}}$ was already introduced in (\ref{def-A0}),  
and the ``mixing-induced'' CP asymmetry is given by 
\begin{equation}\label{def-mix} 
{\cal A}^{\mbox{{\scriptsize mix}}}_{\mbox{{\scriptsize CP}}} 
=\frac{2\,\mbox{Im}\,\xi}{1+|\xi|^2}\,, 
\end{equation} 
with 
\begin{equation}\label{xi-def} 
\xi=e^{-i\phi}\left[\frac{1+\sum_k v_0^{(k)}e^{i\Delta_0^{(k)}} 
e^{-i\Phi_0^{(k)}}+\sum_j v_1^{(j)}e^{i\Delta_1^{(j)}} 
e^{-i\Phi_1^{(j)}}}{1+\sum_k v_0^{(k)}e^{i\Delta_0^{(k)}} 
e^{+i\Phi_0^{(k)}}+\sum_j v_1^{(j)}e^{i\Delta_1^{(j)}} 
e^{+i\Phi_1^{(j)}}}\right]. 
\end{equation} 
Here (\ref{ampl-NPd}) was used to parametrize the decay amplitudes.  
The CP-violating phase $\phi$ is given by $\phi=\phi_{\rm M}+\phi_{K}$,  
where $\phi_{\rm M}$ and $\phi_K$ are the weak $B^0_d$--$\overline{B^0_d}$ 
and $K^0$--$\overline{K^0}$ mixing phases, respectively.  
 
As in our analysis of the $B\to J/\psi K$ system \cite{FM}, it is useful  
to introduce the following combinations of the observables  
(\ref{def-AP}) and (\ref{def-A0}): 
\begin{equation} 
{\cal S}\equiv\frac{1}{2}\left[{\cal A}_{\rm CP}^{{\rm dir}}+{\cal A}_{\rm   
CP}^{(+)} 
\right],\quad 
{\cal D}\equiv\frac{1}{2}\left[{\cal A}_{\rm CP}^{{\rm dir}}-{\cal A}_{\rm   
CP}^{(+)} 
\right]. 
\end{equation} 
The interesting feature of these combinations is that ${\cal S}$ is 
governed by the $I=0$ pieces of the $B\to \phi K$ amplitudes, whereas 
${\cal D}$ and ${\cal B}$ are proportional to the $I=1$ amplitudes.  
 
As the general expressions are very complicated and not very instructive,  
let us focus on the simplified case where the new-physics contributions to  
the $I=0$ and $I=1$ sectors involve either the same weak or strong phases: 
\begin{eqnarray} 
A(B^+\to \phi K^+)&=&{\cal A}_{\rm SM}^{(0)} 
\left[1+v_0 e^{i\Delta_0} e^{i\Phi_0}-v_1 e^{i\Delta_1}e^{i\Phi_1} 
\right]\label{ampl-NPp-simple}\\ 
A(B^0_d\to \phi K^0)&=&{\cal A}_{\rm SM}^{(0)} 
\left[1+v_0 e^{i\Delta_0} e^{i\Phi_0}+v_1 e^{i\Delta_1} e^{i\Phi_1} 
\right].\label{ampl-NPd-simple}  
\end{eqnarray} 
Then we obtain 
\begin{eqnarray} 
{\cal S}&=&-2\left[\frac{a c-b d\,v_1^2}{c^2-d^2\,v_1^2}\right]= 
-2\,v_0\left[\frac{\sin\Delta_0\sin\Phi_0}{1+ 
2\,v_0\cos\Delta_0\cos\Phi_0+v_0^2}\right]+{\cal O}(v_1^2)\label{S-expr}\\ 
{\cal D}&=&
-2\,v_1\left[\frac{b c - a d}{c^2-d^2\,v_1^2}\right]\label{D-expr}\\ 
{\cal B}&=&v_1\,\frac{d}{c}=2\,v_1\left[\frac{\cos\Delta_1\cos\Phi_1+ 
v_0\cos(\Delta_0-\Delta_1)\cos(\Phi_0-\Phi_1)}{1+ 
2\,v_0\cos\Delta_0\cos\Phi_0+v_0^2+v_1^2}\right],\label{B-expr} 
\end{eqnarray} 
with 
\begin{eqnarray} 
a&=&v_0\sin\Delta_0\sin\Phi_0\\ 
b&=&\sin\Delta_1\sin\Phi_1+v_0\sin(\Delta_0-\Delta_1)\sin(\Phi_0-\Phi_1)\\ 
c&=&1+2\,v_0\cos\Delta_0\cos\Phi_0+v_0^2+v_1^2\\ 
d&=&2\left[\cos\Delta_1\cos\Phi_1+v_0\cos(\Delta_0-\Delta_1) 
\cos(\Phi_0-\Phi_1)\right].  
\end{eqnarray} 
Let us finally give the  expression for the mixing-induced CP asymmetry  
(\ref{def-mix}): 
\begin{equation}\label{Amix-calc} 
{\cal A}^{\mbox{{\scriptsize mix}}}_{\mbox{{\scriptsize CP}}}=-\sin\phi- 
2\,\frac{z}{n}, 
\end{equation} 
where 
\begin{eqnarray} 
z&=&\left[v_0\cos\Delta_0\sin\Phi_0+v_1\cos\Delta_1\sin\Phi_1\right]\cos\phi 
\nonumber\\ 
&&+v_0v_1\left[\cos(\phi+\Phi_1)\sin\Phi_0+\cos(\phi+\Phi_0)\sin\Phi_1\right] 
\cos(\Delta_0-\Delta_1)\nonumber\\ 
&&+v_0^2\cos(\phi+\Phi_0)\sin\Phi_0+v_1^2\cos(\phi+\Phi_1)\sin\Phi_1, 
\end{eqnarray} 
and 
\begin{eqnarray} 
n&=&1+2\,v_0\cos\Delta_0\cos\Phi_0+2\,v_1\cos\Delta_1\cos\Phi_1\nonumber\\ 
&&+2\,v_0v_1\cos(\Delta_0-\Delta_1)\cos(\Phi_0-\Phi_1)+v_0^2+v_1^2. 
\end{eqnarray}

\section{Discussion}\label{sec:disc} 
Because of large uncertainties, the data on $B\to \phi K$ decays that  
are available at present from the Belle and CLEO collaborations (see  
(\ref{EXP})) do not allow us to speculate on new physics. However, the  
experimental situation should significantly improve in the next couple of  
years. In this section, we discuss various patterns of the observables  
${\cal S}$, ${\cal D}$ and ${\cal B}$ introduced above that may shed  
light both on new physics and on the $B\to \phi K$ hadron dynamics.  
 
An interesting probe for new physics is of course also provided by the  
mixing-induced CP asymmetry of the $B_d\to \phi K_{\rm S}$ channel, which  
can be compared with the one of the ``gold-plated'' mode  
$B_d\to J/\psi K_{\rm S}$ \cite{BphiK-NP}. Using the expression for  
${\cal A}^{\mbox{{\scriptsize mix}}}_{\mbox{{\scriptsize CP}}} 
(B_d\to J/\psi K_{\rm S})$ given in \cite{FM}, the parametrization for  
${\cal A}^{\mbox{{\scriptsize mix}}}_{\mbox{{\scriptsize CP}}} 
(B_d\to \phi K_{\rm S})$ given in (\ref{Amix-calc}), and the hierarchy  
arising in (\ref{ampl-hier1}), we obtain 
\begin{equation}\label{mix-diff1} 
{\cal A}^{\mbox{{\scriptsize mix}}}_{\mbox{{\scriptsize CP}}} 
(B_d\to \phi K_{\rm S})- 
{\cal A}^{\mbox{{\scriptsize mix}}}_{\mbox{{\scriptsize CP}}} 
(B_d\to J/\psi K_{\rm S})=\underbrace{{\cal O}(1)}_{{\rm NP}_{I=0}} + 
\underbrace{{\cal O}(\overline{\lambda})}_{{\rm NP}_{I=1}} 
+\underbrace{{\cal O}(\overline{\lambda}^2)}_{\rm SM}\, , 
\end{equation} 
where the $\sin\phi$ terms, which may deviate from the Standard-Model  
expectation \cite{AL}, cancel. The contributions entering at the  
$\overline{\lambda}$ and $\overline{\lambda}^2$ levels may also contain  
new-physics effects from $B_d\to J/\psi K_{\rm S}$, whereas the  
${\cal O}(1)$ term would essentially be due to new physics in  
$B_d\to \phi K_{\rm S}$.  
 
In order to disentangle the $I=0$ and $I=1$ isospin sectors, the 
observables ${\cal S}$, ${\cal D}$ and ${\cal B}$ play a key role. 
As can be seen in (\ref{S-expr})--(\ref{B-expr}), ${\cal S}$ provides a  
``smoking-gun'' signal for new-physics contributions to the $I=0$ amplitude,  
whereas ${\cal D}$ and ${\cal B}$ probe new-physics effects in the $I=1$  
sector. Using the hierarchy arising in (\ref{ampl-hier1}), we obtain 
\begin{equation}\label{obs-hier1} 
{\cal S}=\underbrace{{\cal O}(1)}_{{\rm NP}_{I=0}}+ 
\underbrace{{\cal O}(\overline{\lambda}^2)}_{\rm SM},\quad 
{\cal D}=\underbrace{{\cal O}(\overline{\lambda})}_{{\rm NP}_{I=1}} 
+\underbrace{{\cal O}(\overline{\lambda}^2)}_{\rm SM},\quad 
{\cal B}=\underbrace{{\cal O}(\overline{\lambda})}_{{\rm NP}_{I=1}} 
+\underbrace{{\cal O}(\overline{\lambda}^2)}_{\rm SM}, 
\end{equation} 
where the Standard-Model contributions are not under theoretical control.  
If the dynamical suppression of the $I=1$ contributions were larger 
than ${\cal O}(\overline{\lambda})$, ${\cal D}$ and ${\cal B}$ would be  
further suppressed with respect to ${\cal S}$. On the other hand, if  
the rescattering effects described by (\ref{rescatter-u}) were very  
large -- and not small, as assumed in (\ref{obs-hier1}) -- {\it all}  
observables would be of ${\cal O}(1)$. In such a situation, we would not  
only have signals for physics beyond the Standard Model, but also for  
large rescattering processes.  
 
The discussion given above corresponds to the most optimistic scenario 
concerning the generic strength of the new-physics effects in the  
$B\to \phi K$ system. Let us now consider a more pessimistic case,  
where the new-physics contributions are smaller by a factor of  
${\cal O}(\overline{\lambda})$: 
\begin{equation}\label{ampl-hier3} 
A(B\to \phi K)={\cal A}_{\rm SM}^{(0)}\biggl[1+ 
\underbrace{{\cal O}(\overline{\lambda})}_{{\rm NP}_{I=0}}+ 
\underbrace{{\cal O}(\overline{\lambda}^2)}_{{\rm NP}_{I=1}}+ 
\underbrace{{\cal O}(\overline{\lambda}^2)}_{{\rm SM}}\biggr]. 
\end{equation} 
Now the new-physics contributions to the $I=1$ sector can no longer be  
separated from the Standard-Model contributions. However, we  
would still get an interesting pattern for the $B\to \phi K$ observables, 
providing evidence for new physics: whereas (\ref{mix-diff1}) and  
${\cal S}$ would both be sizeable, i.e.\ of ${\cal O}(10\%)$ and within  
reach of the $B$-factories, ${\cal D}$ and ${\cal B}$ would be strongly  
suppressed. However, if these two observables, in addition to  
(\ref{mix-diff1}) and ${\cal S}$, are found to be also at the $10\%$  
level, new physics cannot be distinguished from Standard-Model  
contributions, which could also be enhanced to the $\overline{\lambda}$  
level by large rescattering effects. This would be the most unfortunate 
case for the search of new-physics contributions to the $B\to \phi K$ 
decay amplitudes. An analogous discussion of the $B\to J/\psi K$ system 
was given in \cite{FM}.

\section{Summary}\label{sec:concl} 
The decays $B^\pm\to \phi K^\pm$ and $B_d\to \phi K_{\rm S}$ offer 
an interesting probe to search for physics beyond the Standard Model. 
Using estimates borrowed from effective field theory, we have 
explored the impact of new physics on the $B\to \phi K$ system in a  
model-independent manner, and have derived parametrizations of the  
corresponding decay amplitudes, which rely only on the $SU(2)$ isospin  
symmetry of strong interactions. We have introduced -- in addition to  
the mixing-induced CP asymmetry of the $B_d\to \phi K_{\rm S}$ channel --  
a set of three observables, providing the full picture of possible  
new-physics effects in $B\to\phi K$ decays. In particular, these observables  
allow us to separate the new-physics contributions to the $I=0$ and $I=1$  
isospin sectors, and offer, moreover, valuable insights into hadron dynamics.  
Imposing dynamical hierarchies of amplitudes, we have discussed various  
patterns of these observables, including also scenarios corresponding to  
small and large rescattering effects. We find that $B\to\phi K$ decays  
may provide, in general, powerful ``smoking-gun'' signals for new physics.  
However, there is also an unfortunate case, where such effects cannot be  
distinguished from those of the Standard Model. We strongly encourage our  
experimental colleagues to focus not only on the measurement of the  
mixing-induced CP asymmetries in $B_d\to \phi K_{\rm S}$ and  
$B_d\to J/\psi K_{\rm S}$, but also on the observables ${\cal S}$,  
${\cal D}$ and ${\cal B}$. Hopefully, these will yield evidence for physics  
beyond the Standard Model.

\section*{Acknowledgements} 
The work of T.M. is supported by the DFG Graduiertenkolleg 
``Elementarteilchenphysik an Beschleunigern'', by the 
DFG Forschergruppe ``Quantenfeldtheorie, Computeralgebra und Monte   
Carlo Simulationen'', and by the 
``Ministerium f\"ur Bildung und Forschung'' (bmb+f).

\end{document}